\documentclass[aps,prl,superscriptaddress,twocolumn,showpacs]{revtex4-1}

\usepackage{graphicx}
\usepackage{bm}
\usepackage{subfigure}
\usepackage{amsmath,amsfonts}
\usepackage{color}
\usepackage[normalem]{ulem}

\begin{document}

\title{Matter-Wave Interference versus Spontaneous Pattern Formation in Spinor Bose-Einstein Condensate}

\author{Marcin Witkowski}
\affiliation{Institute of Physics, Faculty of Physics, Astronomy and Informatics, Nicolaus Copernicus University, Grudzi\c{a}dzka 5, PL-87-100 Toru\'n, Poland,}
\affiliation{Institute of Physics, University of Opole, Oleska 48, PL-45-052 Opole, Poland,}
\author{Rafa\l{} Gartman}
\affiliation{Institute of Physics, Faculty of Physics, Astronomy and Informatics, Nicolaus Copernicus University, Grudzi\c{a}dzka 5, PL-87-100 Toru\'n, Poland,}
\author{Bart\l{}omiej Nag\'orny}
\affiliation{Institute of Physics, Faculty of Physics, Astronomy and Informatics, Nicolaus Copernicus University, Grudzi\c{a}dzka 5, PL-87-100 Toru\'n, Poland,}
\author{Marcin Piotrowski}
\affiliation{Institute of Physics, Faculty of Physics, Astronomy and Informatics, Nicolaus Copernicus University, Grudzi\c{a}dzka 5, PL-87-100 Toru\'n, Poland,}
\affiliation{Institute of Physics, Jagiellonian University, Reymonta 4, PL-30-059 Krak\'ow, Poland,}
\author{Marcin P\l{}odzie\'{n}}
\affiliation{Institute of Physics, Jagiellonian University, Reymonta 4, PL-30-059 Krak\'ow, Poland,}
\author{Krzysztof Sacha	}
\affiliation{Institute of Physics, Jagiellonian University, Reymonta 4, PL-30-059 Krak\'ow, Poland,}
\author{Jacek Szczepkowski}
\affiliation{Institute of Physics, Polish Academy of Sciences, al. Lotnik\'ow 32-46, PL-02-668 Warszawa, Poland,}
\author{Jerzy Zachorowski}
\affiliation{Institute of Physics, Jagiellonian University, Reymonta 4, PL-30-059 Krak\'ow, Poland,}
\author{Micha\l{} Zawada}
\affiliation{Institute of Physics, Faculty of Physics, Astronomy and Informatics, Nicolaus Copernicus University, Grudzi\c{a}dzka 5, PL-87-100 Toru\'n, Poland,}
\email[]{zawada@fizyka.umk.pl}
\author{Wojciech Gawlik}
\affiliation{Institute of Physics, Jagiellonian University, Reymonta 4, PL-30-059 Krak\'ow, Poland,}

\date{\today}

\begin{abstract}

We describe effects of matter-wave interference of spinor states in the $^{87}$Rb Bose-Einstein condensate. The components of the $F=2$ manifold are populated by forced Majorana transitions and then fall freely due to gravity in an applied magnetic field. Weak inhomogeneities of the magnetic field, present in the experiment, impose relative velocities onto different $m_F$ components, which show up as interference patterns upon measurement of atomic density distributions with a Stern-Gerlach imaging method. We show that interference effects may appear in experiments even if gradients of the magnetic field components are eliminated but higher order inhomogeneity is present and the duration of the interaction is long enough.
In particular, we show that the resulting matter-wave interference patterns can mimic spontaneous pattern formation in the quantum gas. 

\end{abstract}


\maketitle

Atom interferometry is a well established technique which allows sensitive measurements of many physical quantities. 
Important examples include atomic clocks and frequency standards \cite{interferometry1,interferometry2}, precision gravitational measurements \cite{interferometry3,interferometry4}, and measurements of fundamental physical constants \cite{interferometry5,interferometry6}. One crucial parameter that determines the interference structure is the phase difference of interfering components.  As discussed below, in spinor condensates this phase difference can be efficiently controlled by a magnetic field which paves the way towards magnetometric applications (earlier work on cold-atom magnetometry has been recently reviewed in Ref. \cite{magnetometry1}).

In this work we report on the application of spinor Bose-Einstein condensate (BEC) to atom-wave interferometry and demonstrate its sensitivity to very weak magnetic fields. We indicate that even in meticulously performed experiments the residual magnetic fields and their inhomogeneity can create interference patterns of atomic density distributions which seriously affect the investigated atomic samples.
Apart from possible practical applications, like in magnetometry and/or imaging, we show that the phase shifts of individual components of the spinor condensate may often mimic the effects of other interactions. One example of such effects is the spontaneous pattern formation resulting from the interaction of spinor components, which has been intensively studied recently \cite{sengstock,raman,stamper_ueda}.

Below we present results of our studies with the spinor BEC of $^{87}$Rb atoms ($F=2$) which expand the field of atomic interferometry to spinor condensates and demonstrate very high sensitivity of the distributions of atomic densities to weak magnetic fields. The observations are interpreted  theoretically in terms of the matter-wave interference.
We also investigate theoretically matter-wave interference in the case of trapped atoms when there is no first-order gradient but a curvature of the axial magnetic field component. The analysis shows that the matter-wave interference is a very general phenomenon which can occur in many specific experimental situations.

Most of the previous studies of spinor condensates were conducted with optical dipole traps which store atoms with all spin components rather than just the low magnetic-field seeking spin states of magnetic traps \cite{stamper_ueda}. In the described experiment, however, the condensate is formed by $^ {87}$Rb atoms in a single  magnetic component of their ground state contained in the magnetic trap (MT). An additional difference between the present and previous studies is that, in contrast to the {\emph{•} in situ} experiments with high-density samples, we work with an expanding condensate where the atomic densities are sufficiently low that the atomic interaction can be neglected.

In the experiment, we collect up to $10^6$ $^{87}$Rb atoms in the $|F = 2, m_F = 2\rangle_x$ state in the Ioffe-Pritchard QUIC-like MT \cite{APPA.113.691}.
The axis of the elongated condensate is aligned horizontally, along the $x$ axis.
The sample is analyzed in the horizontal radial direction by an absorption imaging with the resonant probe beam in the $y$ direction.
A system of coils compensates the ambient magnetic fields to below 25~mG and  imposes a weak field during gravitational free fall of the BEC released from the trap. Additional two coils control the inhomogeneity of the fields in two orthogonal directions down to about 7~mG/cm. The magnetic field created by the coils is calibrated by observing an expansion of the atomic cloud in the field of six counter-propagating molasses beams of equal intensities.
The electronically controlled turning off the MT and an extra pair of the Helmholtz coils which alter the decay of the magnetic field on the axis of the Ioffe coil produce a zero transition of one of the components of the magnetic field with a variable speed. At the crossing, the Majorana transitions \cite{Majorana} mix the $|m_F\rangle_x$ sublevels and populate various magnetic states of the atoms in a way similar to that reported in \cite{PhysRevA.73.013624}. The approaching of the total magnetic field to the zero value is crucial for the experiment. Its details are described elsewhere \cite{Witkowskitobe}.

\begin{figure}
\includegraphics[width=8cm]{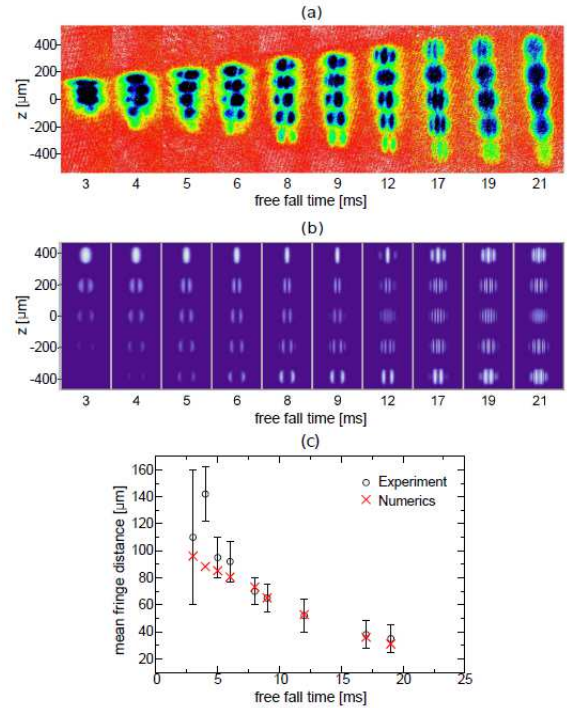}
\caption{(color online) Spinor condensates after a free fall of indicated time and separated vertically by the SG force pulse. Spatial separation reveals the interference pattern in thr form of vertical fringes in  the five spinor components. (a) Experimental results. Different vertical splittings reflect different values of the SG gradient at different distances from the trap center. (b) Results of numerical simulations as described in the text. (c) Dependence of the mean fringe distance in the $m_F=2$ component on the free fall time.}
\label{fig:fringes}
\end{figure}

After mixing of the $|m_F\rangle_x$ states the MT field is adiabatically replaced by a weak, inhomogeneous magnetic field $\vec B_d$. The spinor BECs fall freely due to gravity in this field. The gradient of the $x$ component of $\vec B_d$ leads to new relative velocities of the $|m_F\rangle_{x}$ components, which are, however, too small to spatially separate the components. For typical values of the applied gradients of tens of mG/cm and time of flight in the presence of the field of 2 to 20~ms, the resulting velocities are on the order of 50 $\mu$m/s. Such low velocities are sufficient for the interference effects discussed below.
After the time of flight, the spinor BECs are spatially separated by a vertical Stern-Gerlach (SG) force acting for 1--2~ms. The population distribution among individual  $|m_F\rangle_z$ components is recorded 3~ms later. The strong vertical SG force projects atoms onto a new quantization axis. Atoms belonging to different spinor components, moving with different velocities can be projected onto the same $|m_F\rangle_z$ states which results in the interference pattern, see Fig.~\ref{fig:fringes}.

In the modeling we assume the particles are initially in the zero momentum state $\psi(\vec {r})$ and in a combination of $|m_F\rangle_x$ states which we choose to be $|m_F=2\rangle_z$, as a generic example. All experimental absorption images are obtained in the $XZ$ plane; hence, we restrict the discussion to the two-dimensional (2D) model.

The presence of a slightly inhomogeneous magnetic field directed along the $x$-axis, $\vec B_d=(B_{d0}+x\frac{\partial B_d}{\partial x})\hat x$, for the time from $t=0$ to $t_I$ results in a force different for different spin components $m_F$. The momentum transfer also depends on $m_F$ and each component acquires a phase proportional to $m_FB_{d0}t$. In a simple analytic model at $t=t_I$ each $|m_F\rangle_x$ component is of the form
\begin{equation}
\psi_{k_{m_F}}(x,z)\;|m_F\rangle_x=e^{im_F(kx+\phi)}\;|m_F\rangle_x
\label{psikmf}
\end{equation}
with $k=g_F \mu_B \frac{\partial B_d}{\partial x}t_I/\hbar$ and $\phi=g_F \mu_B B_{d0}t_I/\hbar$.

The density structures are visualized by projecting $|\varPsi(t_I)\rangle$ on different $|m_F\rangle_z$ states to obtain
\begin{equation}
\rho_{m_F}(x,z)=\left|_z\langle m_F|\varPsi(t_I)\rangle\right|^2.
\label{rho}
\end{equation}
Each $\rho_{m_F}$ reveals an interference pattern resulting from the interaction of an atom with the inhomogeneous field $\vec B_d$.
Note that by changing $B_{d0}$, or changing slightly the interaction time $t_I$, we can shift the interference pattern along the $x$ direction, since the phase $\phi$ in Eq. (\ref{psikmf}) is a common parameter of translation for all plane waves.

In the full numerical simulation we assume the initial Thomas-Fermi (TF) profile wave packet of the form $\psi(x,z)=\sqrt{\frac{2}{\pi}}\sqrt{R^2-x^2-z^2}/R^2$ with  $R=60 $~$\mu$m.
The non-linear term in the Gross-Pitaevskii (GP) equation \cite{Gross,Pitaevskii} is negligible because after turning off the MT, atomic gas expands and particle interactions become quickly negligible.
Thus, we integrate the Schr\"odinger equation with the magnetic interaction term $g_F\mu_B\hat{\vec{F}}\cdot\vec{B}$ where
$\hat{\vec{F}}$ stands for the spin operator \cite{ueda}. Then, the SG pulse and further evolution are included. To avoid high momenta in the numerical simulation, slightly different parameters of the SG field were chosen, i.e., 500~mG/cm acting for 20~ms. In that way, the fast spatial separation of different $m_F$ components is substituted for longer evolution but with smaller momenta.

\begin{figure}[hbt]
 \includegraphics[width=0.9\columnwidth]{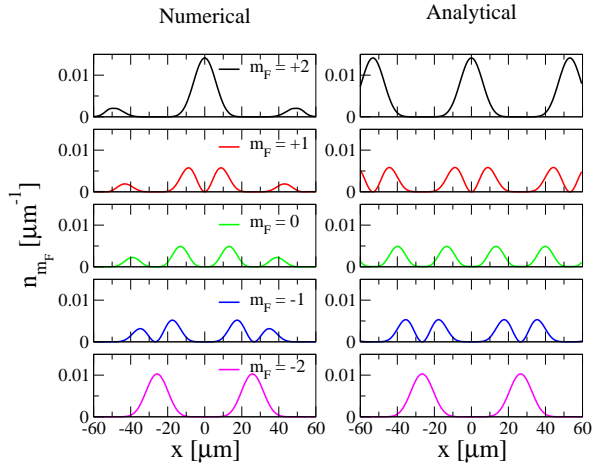}
 \caption{(color online) Integrated probability densities $n_{m_F}(x)$ at $t=t_I$. The left panels correspond to the numerical simulation and the right ones to the results of a simple analytical model.
\label{fig:theor}}
\end{figure}

Figure~\ref{fig:theor} shows integrated densities $n_{m_F}(x)=\int \rho_{m_F}(x,z)dz$, where $\rho_{m_F}(x,z)$ is given in Eq. (\ref{rho}), at $t=t_I$ compared to the prediction of the simple analytic model. We have chosen $B_{d0}=0$,
$\frac{\partial B_{d}}{\partial x}=22.3$~mG/cm and $t_I=12$~ms.
In the analytical model the spatial degrees of freedom are described by plane waves, consequently there is no overall change in the corresponding density profiles. However, structures of the interferences patterns are perfectly the same as in the case of the numerical approach.

The results of the numerical simulation are in a good agreement with the experimental data (Fig.~\ref{fig:fringes}). In both cases we can see clear antisymmetry between the interference fringes for  $m_F$ and $-m_F$ components. Vertical integration of the images in Figs.~\ref{fig:fringes} (a), (b) and 2 yields structureless distributions reconstructing the initial TF profiles. Because of the limited resolution of our detection system, the theoretically predicted interference patterns reveal more fine structures than in the experiment.
Moreover, even a very small force acting in the $y$ direction can rotate the interference pattern so that the fine density structures become blurred because the direction of the absorption imaging is not parallel to the planes of the interference fringes. 
In contrast to numerical simulations which were performed for a constant SG gradient, the experimental patterns expand vertically in time, since the SG field in our trap has different gradients at different distances from the trap center.

\begin{figure}
    \includegraphics[width=0.7\columnwidth]{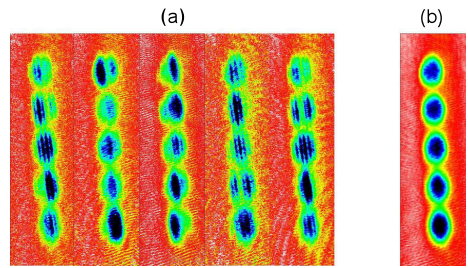}
    \caption{(color online) (a) Five successive realizations of the interference starting from the same experimental conditions. (b) The image averaged over 72 realizations.}
    \label{fig:mean}
\end{figure}

Successive realizations of the experiment (Fig. \ref{fig:mean}(a)) show that the horizontal positions of the interference fringes vary since each experiment is realized in about 1--2~min.~intervals and the phase of the fringes strongly depends on the $B_{d0}$ value. Under conditions of our experiment the phase change of $\pi$ corresponds to $\Delta B_{d0} \approx 10^{-4}$ G, which is close to the long-term stability limit of our power supplies.
Because of random phases in each of the realizations, after averaging over 72 measurements, the interference structures disappear in regular TF envelopes of each BEC component (Fig. \ref{fig:mean}(b)).

It is very difficult to eliminate all external magnetic fields in a laboratory. Consequently, if there is a very weak inhomogeneous magnetic field but the time of its interaction with atoms is long, the interference effects can arise. If the strength of the field changes slightly from one experimental realization to another, or the interaction time is not repeated with the precision much better than the Larmor period, the interference fringes can move by more than their spatial period and the experimental data appear as if there was spontaneous pattern formation.

Even if the linear gradient is compensated, the higher order inhomogeneities may be still present in experimental realizations. We have verified that, a small, second order inhomogeneity can result in interference effects even in the presence of the trap and particle interactions.
To this end we integrate the 1D GP equation, $i\hbar\partial_t\psi_{m_F}=\partial{\cal H}/\partial\psi_{m_F}^*$, corresponding to the energy density \cite{ueda}
\begin{eqnarray}
{\cal H}&=&\sum_{m_F}\psi_{m_F}^*\left[-\frac{\hbar^2\partial_x^2}{2m}+\frac{m\omega^2x^2}{2}
-p(x)\;m_F \right.
\cr
&+& \left.q(x)\;m_F^2\right]\psi_{m_F} +\frac{c_0}{2}n^2(z)+\frac{c_1}{2}|\vec F|^2+\frac{c_2}{2}|\Theta|^2,
\label{gpe}
\end{eqnarray}
where $p(x)$ and $q(x)$ stand for the linear and quadratic Zeeman energy, respectively. They depend on $x$ because we assume the non-vanishing curvature of the axial magnetic field, $\vec B=\left(B_0+\alpha x^2/2\right)\hat x$.

\begin{figure}[hbt]
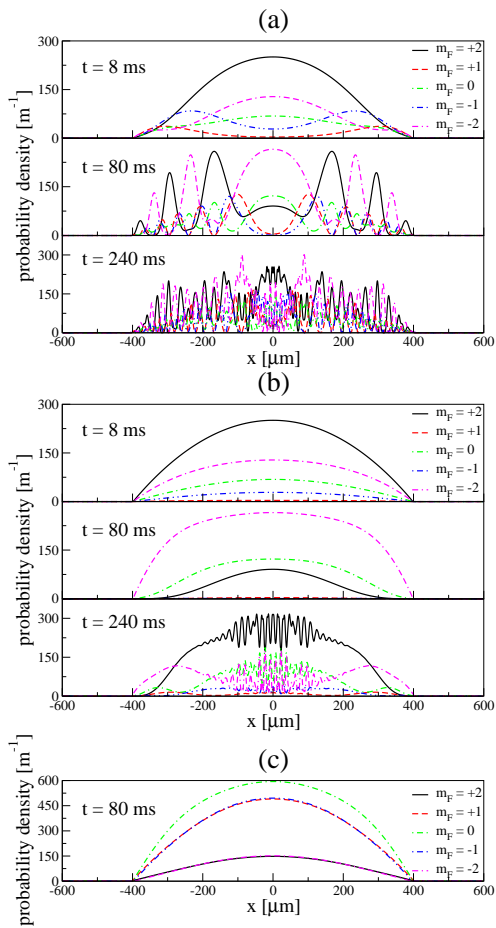

\includegraphics[width=0.75\columnwidth]{B.1.10.spin.eps}
\includegraphics[width=0.75\columnwidth]{B.1.10.spin_no.curv.eps}
\includegraphics[width=0.75\columnwidth]{B.1.10_spin_z.eps}
\caption{(color online) (a) Atomic densities after evolution in the presence of the harmonic trap and the inhomogeneous magnetic field with $\alpha=60$~mG/cm$^2$ for different times $t$ as indicated in the panels. Different spin components $|m_F\rangle_z$ are plotted with different lines. (b) The same as in (a) but in the absence of the field inhomogeneity, i.e., for $\alpha=0$. In (c) we show similar data as in the middle panel of (a) but the densities correspond to different $|m_F\rangle_x$ components, i.e., we project the spin state on the $x$ axis.
\label{fig:seng}}
\end{figure}

In our simulations of the effect of the second-order magnetic field inhomogeneity on the matter-wave interference we apply  parameters corresponding to the experiment reported in Ref.~\cite{sengstock}, i.e., $B_0=1.1$~G, the curvature $\alpha=60$~mG/cm$^2$,  $\omega/2\pi=0.8$~Hz, the interaction coefficients in 1D $c_0=0.116\sqrt{\hbar^3\omega/m}$, $c_1=1.64 \times 10^{-2}c_0$, $c_2=-1.72 \times 10^{-3}c_0$, and total number of atoms $\int n(x) dx=\int dx \sum_{m_F}|\psi_{m_F}|^2=2.1 \times 10^5$. Initially, atoms are prepared in the state  $|\Psi(0)\rangle=\psi(x)|m_F=2\rangle_z$, where $\psi(x)$ is the TF profile of a single component condensate in the harmonic trap.

The initial spin state $|m_F=2\rangle_z$ is a superposition of different $|m_F\rangle_x$ states.
The presence of the inhomogeneous field leads to different trapping potentials for different $|m_F\rangle_x$. Indeed, the linear Zeeman term makes the harmonic trap slightly more shallow or more steep depending on $m_F$ (the frequency change corresponding to the considered conditions is $0.038 m_F\omega$). Consequently, atomic wavepackets corresponding to different signs of $m_F$ start breathing out of phase. Such a relative motion results in interference patterns if at the end of the temporal evolution $|\Psi(t)\rangle$ is projected on different  $|m_F\rangle_z$ states, see Fig.~\ref{fig:seng}(a).
However, if $|\Psi(t)\rangle$ is projected on the $|m_F\rangle_x$ states no interference fringes emerge because there is no mixing of the differently breathing wavepackets. It is illustrated in Fig.~\ref{fig:seng}(c) where we show similar data as in Fig.~\ref{fig:seng}(a) but with the densities corresponding to different $|m_F\rangle_x$ components. The projection on the $|m_F\rangle_x$ components was used in experiment  \cite{sengstock} where the SG field separated spatially the $|m_F\rangle_x$ components, rather than the $|m_F\rangle_z$ ones.

Figure~\ref{fig:seng}(a)
shows that the first signature of the interference effects is visible already at $t\approx 8$~ms. At $t=80$~ms we can see well developed interference patterns while for $t\ge 240$~ms the density profiles become irregular. The irregular profiles appear due to the presence of the spin dependent interactions. If we turn off these interactions, we observe the regular patterns for times much greater than 240~ms.
For a given $B_0$ and $t$, the interference pattern is the same in different experimental realizations, provided the evolution time $t$ does not change more than a small fraction of the Larmor precession period and $B_0$ does not fluctuate more than, e.g., $10^{-5}$~G for $t=80$~ms. Otherwise one observes random shifts of the pattern.
The presence of the curvature changes dramatically the system behavior. For comparison, in Fig.~\ref{fig:seng}(b) we present the results for $B_0=1.1$~G but with $\alpha=0$ which are very different from the case shown in Fig.~\ref{fig:seng}(a) where the curvature is included. That is, there is no modulation of the atomic densities in Fig.~\ref{fig:seng}(b) until the dynamical instability becomes visible.

In this report we have analyzed the matter-wave interference of an expanding $^{87}$Rb spinor BEC. The experimental observations agree very well with theoretical simulations and demonstrate that such BECs are extremely sensitive to magnetic field inhomogeneity. 
In addition to the modeling of the experiment described above, we have analyzed the case of higher-order inhomogeneities and applied
 the developed approach to model matter-wave interference in the case when there is no first-order gradient but a curvature of the axial magnetic field component. In the analyzed situation, very well pronounced interference patterns have been obtained.
This study shows, that even very weak magnetic inhomogeneities, difficult to avoid in experiments, result in interference patterns which may mimic the patterns attributed to dynamical instabilities. The matter-wave interference appears thus to be a very general phenomenon, possible to occur in a wide range of experimental conditions. In particular, since it  may occur independently of other interactions in spinor BECs, the fact can not be excluded that in many experiments where pattern formation is attributed to dynamical instability, the atom-wave interference plays an important role. 
We have demonstrated that the interference pattern created by an inhomogeneous magnetic field, with a gradient of 30 mG/cm, for example, acting for about 10 ms will be noticeably shifted if the magnetic field changes by $10^{-4}$ G in different experimental realizations. The described results are thus interesting for precision measurements of magnetic fields. The detailed analysis of such applications goes, however, beyond the scope of the present work and will be published elsewhere ~\cite{Witkowskitobe}

\begin{acknowledgments}
The authors are grateful to M.~Matuszewski, D.~Stamper-Kurn, M.~Gajda, A.~Szczepkowicz, and K.~Rz\c{a}\.zewski for numerous discussions.
This work has been performed in the National Laboratory of AMO Physics in Torun and partially supported by the Polish grants no. N N202 230237, DEC-2011/01/N/ST2/00418 (M. P\l{}odzie\'{n}), DEC-2012/04/A/ST2/00088 (KS), and the TEAM Project of the FNP.
\end{acknowledgments}

\end{document}